%% file: main.tex
\documentclass{article}
\input{packages}

\title{NuGraph2 with Context--Aware Inputs: Physics--Inspired Improvements in Semantic Segmentation}
\input{authors}
\date{\today}

\begin{document}

\maketitle

\begin{abstract}
Graph neural networks have recently shown strong promise for event reconstruction tasks in Liquid Argon Time Projection Chambers, yet their performance remains limited for underrepresented classes of particles, such as Michel electrons. In this work, we investigate physics-informed strategies to improve semantic segmentation within the NuGraph2 architecture. We explore three complementary approaches: (i) enriching the input representation with context-aware features derived from detector geometry and track continuity, (ii) introducing auxiliary decoders to capture class-level correlations, and (iii) incorporating energy-based regularization terms motivated by Michel electron energy distributions. Experiments on MicroBooNE public datasets show that physics-inspired feature augmentation yields the largest gains, particularly boosting Michel electron precision and recall by disentangling overlapping latent space regions. In contrast, auxiliary decoders and energy-regularization terms provided limited improvements, partly due to the hit-level nature of NuGraph2, which lacks explicit particle- or event-level representations. Our findings highlight that embedding physics context directly into node-level inputs is more effective than imposing task-specific auxiliary losses, and suggest that future hierarchical architectures such as NuGraph3, with explicit particle- and event-level reasoning, will provide a more natural setting for advanced decoders and physics-based regularization. The code for this work is publicly available on Github at \url{https://github.com/vitorgrizzi/nugraph_phys/tree/main_phys}.
\end{abstract}

\section{Introduction}
Graph neural networks (GNNs) have emerged as powerful tools in particle physics for analyzing detector data, largely because they can natively represent physical relationships in irregular data \cite{bronstein2017geometric, cao2020comprehensive, atz2021geometric}. Unlike traditional image or sequence models, GNNs allow high-level physical features to be attached to measured data and analyzed by the model \cite{thais2022graph}. Recent work has leveraged this flexibility to inject domain knowledge at various stages of the learning pipeline \cite{dash2022review}. Among these approaches, three pertinent strategies are: enriching inputs with physics-inspired features, multi-task learning with auxiliary outputs, and physics-informed loss constraints. These approaches have demonstrated improved expressivity, disambiguation of classes in sparse data, and physically plausible predictions in tasks relevant to Liquid Argon Time Projection Chambers (LArTPC) detectors and beyond, providing key inspiration for our method.

One line of research augments the raw graph inputs with engineered features that encode geometric or physical context. By summarizing local structures or known physics, these descriptors can enhance GNN expressivity, especially when data are sparse or ambiguous. For example, Drielsma et al. \cite{drielsma2021clustering} introduce 22 geometric descriptors per node (e.g., a fragment's covariance, principal axes, centroid, etc.) in their Graph Particle Aggregator (GrapPA) to help assemble electromagnetic shower fragments in LArTPC data, where a fragment is fully connected cluster of energy deposits (hits) in the detector. Pairwise relationships are likewise encoded: each edge carries 19 physics-motivated features (such as closest approach distances between fragments) to inform the network which pieces likely belong to the same particle shower. These hand-crafted features provide the network with a notion of shape, orientation, and proximity that would be difficult to learn from raw data alone, leading to significantly improved clustering of particle showers. 

More generally, incorporating domain-specific inputs has proven beneficial in other particle physics ML applications. For instance, by combining 3D convolutional neural networks with physics-inspired statistical summaries of the electromagnetic radiation pattern, Kiesler et al. \cite{kieseler2022calorimetric} demonstrate that incorporating domain knowledge significantly enhances the precision of muon energy estimates. Their results highlight the complementary benefits of deep learning and hand-crafted, physics-driven features in this context. Overall, encoding physically relevant attributes at the input stage guide GNNs to learn representations that respect known structures in the data (e.g., detector geometry or particle shower morphology). This idea motivates our inclusion of domain-inspired graph features to provide structural context beyond the raw hit information.

Another strategy is to design networks that jointly predict multiple targets to exploit known correlations among classes or tasks. In general, multi-task learning has been found to improve performance in physics machine learning by providing internal consistency checks: the network is pressured to find representations useful for several related predictions rather than overfit to one. In high-energy physics, many labels are interdependent (for example, the presence of certain particle types in an event is correlated with the event's topology or origin). Rather than training separate models, a single network with auxiliary decoders for related tasks can share representations and teach the model these inter-class relationships. A notable example is the DUNE Collaboration's Convolutional Visual Network (CVN), which uses a Convolutional Neural Network (CNN) to simultaneously classify neutrino flavor, interaction type, and particle content of each event \cite{abi2020neutrino}. The CVN produces multiple sets of output probabilities for neutrino flavor (electron, muon, tau, or none), for interaction channel (e.g., charged-current vs. neutral-current), and even for the counts of protons, pions, and neutrons in the final state. This multi-head architecture leverages the fact that, for instance, an electron-neutrino charged-current event is likely to contain an outgoing electron and proton, whereas a neutral-current event might produce different secondaries. By learning all these labels together, the network's shared hidden features capture the semantic relationships between flavor and topology, yielding more robust classification than any single-task model. 

Beyond model architecture, an important avenue is incorporating physics domain knowledge into the loss function or training objective. By adding custom loss terms or regularization penalties, one can directly encourage the model to produce physically plausible outputs or to respect known laws~\cite{nabian2020physics, nabian2021efficient,djeumou2022neural}. In other words, the idea is to use the loss function as a steering mechanism, gently constraining the model toward physically consistent behavior without hard-coding it. This concept underlies the growing field of physics-informed neural networks (PINNs), where differential equations or other theoretical constraints are imposed as part of the loss~\cite{raissi2019physics}. By design, such constraints drive the learned model towards the manifold of solutions allowed by fundamental physics, improving generalization and trustworthiness of the predictions \cite{thangamuthu2022unravelling}. Even when exact laws are not known or easily applied, domain-informed loss functions can be crafted to encode experiment-specific expectations. In the context of graph networks, Sharma et al.~\cite{sharma2025dynami} recently demonstrated a physics-informed GNN for multi-body dynamics that explicitly enforces conservation laws. Their model adds terms to the loss to conserve linear and angular momentum in sphere collisions, yielding predictions that obey symmetries and remain stable over long simulations. 

Drawing inspiration from these prior works, our study investigates the impact of injecting physics-domain knowledge into the NuGraph2~\cite{nugraph2} network through tailored input features, auxiliary decoders, and loss regularization terms. 
In a LArTPC, three wire planes ($u$, $v$, $y$), arranged at different wire directions, collect ionization electrons drifting through the liquid argon. In NuGraph2, each detector hit is represented as a node. Delaunay triangulation algorithm is used to connect nodes belonging to the same plane (planar nodes), while a set of nexus nodes serves as inter-plane connection points, enabling information exchange between the different wire planes. 
NuGraph2 labels each detector hit (i.e. a planar node) based on a binary discriminator for signal vs background and a semantic classifier, where classes correspond to five different particle types: Minimum Ionizing Particles (MIPs), Highly Ionizing Particles (HIPs), electro-magnetic showers, Michel electrons, and diffuse activity.
A more detailed introduction to NuGraph2 is in~\cite{explainability}.

One of our main goals is to improve the network's ability to correctly label Michel electron hits, which are harder to classify compared to other classes due to their scarcity or underrepresentation in the dataset. Michel electrons are an important signature in LArTPC neutrino detectors. One the one side, they represent a standard candle for the reconstruction of low-energy showers \cite{MicroBooNE:2017kvv, MicroBooNE:2021nss}, thus offering a tool for energy calibration critical for e.g. sterile neutrino searches or detection of  neutrinos from supernova bursts. On the other hand, they also provide a unique handle for discriminating the muon charge and thus enhancing the sensitivity to determine the mass ordering with atmospheric neutrinos~\cite{Ternes:2019sak}.
While our main focus is on Michel electrons, we also examine how Michel-targeted interventions may affect the classification accuracy of other particle classes. Similar to the NuGraph2 result~\cite{nugraph2}, our work is also based on the MicroBooNE public datasets~\cite{abratenko_2023_8370883,abratenko_2022_7261921}, which include simulated neutrino interactions from the Booster Neutrino Beam on top of off-beam data collected with the MicroBooNE detector~\cite{acciarri2017design}. 
MicroBooNE is a 2.6 $\times$ 2.3 $\times$ 10.4 m time projection chamber immersed in liquid argon, corresponding to an active volume of 85 tonnes. Ionization electrons produced by charged particles traversing the argon medium drift in a constant electric field of 273 V/cm towards the anode which is instrumented with three wire planes at different angles (0, $\pm$60 degrees with respect to the vertical direction) that measure the incoming charge and that, once stereo views are combined, provide 3D imaging capabilities. Wires are spaced at a distance of 3 mm, for a total of over 8,200 wires. Signal from each wire is digitized every 0.5 $\mu$s. MicroBooNE operated from 2015 to 2020, and the collaboration authored over 80 publications to date.
More information on these datasets can be found in~\cite{Cerati:2023rtv}. The code implementation is available on Github at \url{https://github.com/vitorgrizzi/nugraph_phys/tree/main_phys}.

\section{Methodology}
To improve the network's ability to classify underrepresented classes in the dataset (Michel electrons in our case) and to enhance overall model performance, we explore three complementary strategies grounded in physics intuition. These strategies primarily aim at enriching the network's understanding through domain-informed guidance. First (Sec.~\ref{sec:featext}), we augment the node input space with domain-inspired features that explicitly encode non-local and structural context, such as node degree, proximity, and directional coherence. Second (Sec.~\ref{sec:decoders}), we introduce auxiliary decoders designed to teach the network meaningful correlations between semantic classes, leveraging relationships such as the conditional presence of Michel electrons on the existence of MIPs. Third (Sec.~\ref{sec:energyreg}), we incorporate domain-informed regularization terms into the overall loss function to constrain network behavior according to known physical distributions, particularly with respect to energy deposition profiles. These three strategies collectively inject inductive biases that guide the network toward more discriminative latent representations, particularly for underrepresented or ambiguous classes such as Michel electrons. 

\subsection{Feature Extension}
\label{sec:featext}
Recent work has revealed that standard GNNs, particularly those based on message passing, are inherently limited in their ability to capture graph structure. In particular, they are proved to be no more expressive than the 1-Weisfeiler-Leman graph isomorphism test and consequently fail to distinguish graph substructures or encode higher-order relational context among nodes \cite{bouritsas2022improving}. This means that GNNs are generally unaware of the structural roles of nodes and cannot capture meaningful patterns such as cycles, cliques, or neighborhood asymmetries, which are often critical in physics-based or scientific domains.

To mitigate the limitations of GNNs in learning contextual information from both the structural roles of nodes and the topological properties of the graph, we engineered features that encode each node's relational and structural context within the graph. These features, derived from domain knowledge, include quantities such as node degree, local geometric distances, and neighbor-relative differences in $time \ vs \ wire$ space. By doing so, we aim to inject inductive biases that GNNs would otherwise struggle to learn from data alone, thereby enhancing the model's expressivity and its ability to resolve class ambiguities, especially in challenging cases such as Michel electron identification.

\begin{figure}
    \centering
    \includegraphics[scale=0.38, trim=0cm 1cm 0cm 0cm, clip]{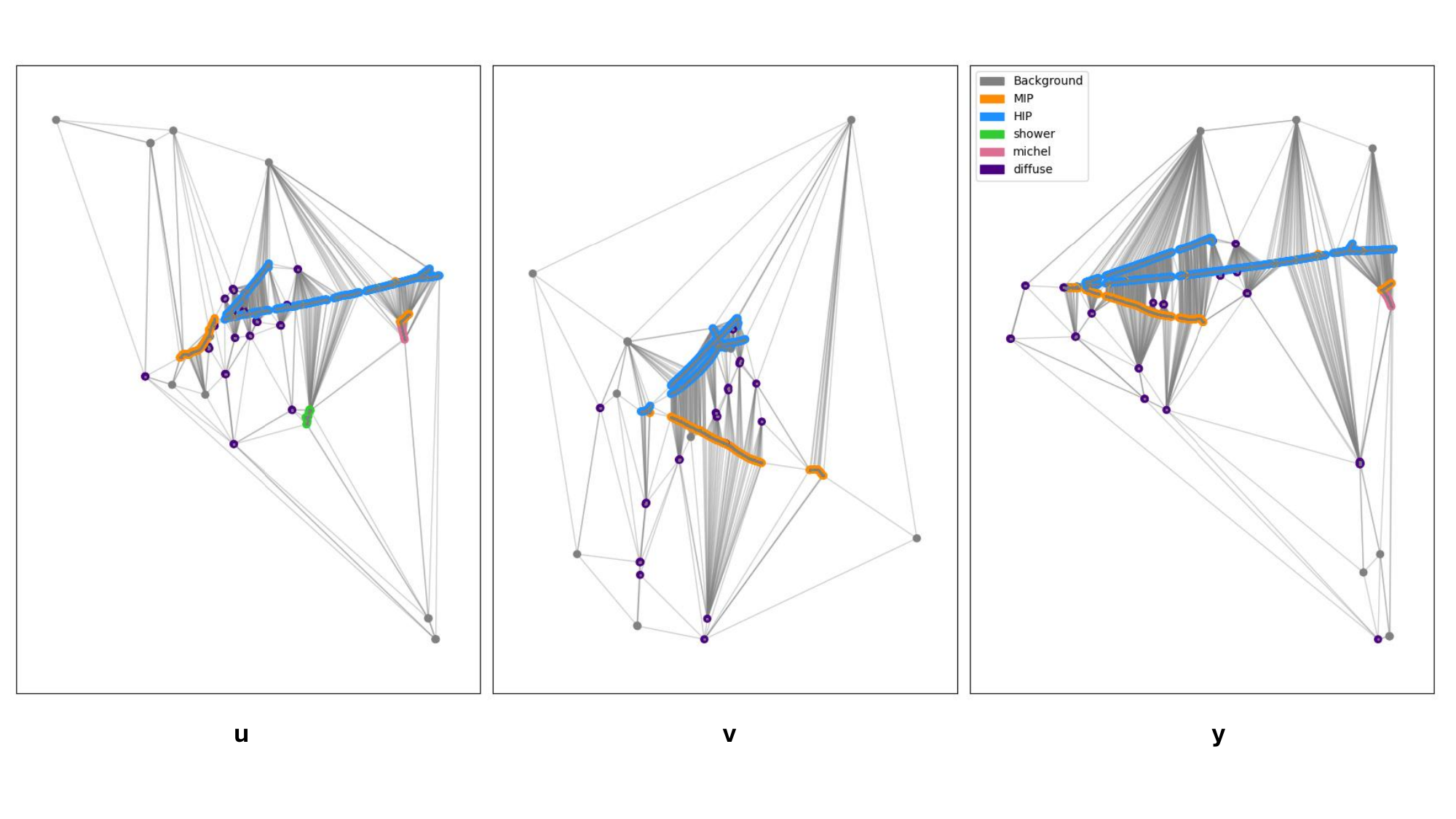}
    \caption{Node connectivity in the  $time \ vs \ wire$ plane for the three different planes $u$, $v$, and $y$. Each node represents a detector hit and edges were constructed using Delaunay triangulation algorithm. Nodes are color-coded according to their true semantic label. The graph exhibits linear tracks characteristic of HIPs and MIPs, and dispersed nodes with high connectivity acting as ``hubs``. Both patterns served as inspiration for augmenting the input features.} 
    \label{fig:edges}
\end{figure}

To improve the overall performance of the model and incorporate domain insights, we extended the original four input features of NuGraph2 planar nodes, i.e. the hit time, wire, width, and integral. These new features were crafted using specific domain knowledge and encode non-local and structural information not explicitly captured by the original features. The additional features are:
\begin{itemize}
    \item Node degree.
    \item Shortest edge length in the $time \ vs \ wire$ plane.
    \item Double difference between the $wire$ value of the current node and its two nearest neighbors in the $time \ vs \ wire$ plane. This feature is denoted by $\Delta wire$.
    \item Double difference between the $time$ value of the current node and its two nearest neighbors in the $time \ vs \ wire$ plane. This feature is denoted by $\Delta time$.
\end{itemize}
The inspiration to include these features was guided by inspecting the neutrino-interaction event graphs. Figure~\ref{fig:edges} illustrates one such event in the $time \ vs \ wire$ plane for the $u$, $v$, and $y$ detector planes. It shows coherent linear tracks, characteristic of HIPs and MIPs, and nodes with very distinct structural roles in the graph, as indicated by their connectivity and nearest-neighbor distance. These patterns were key to the construction of the additional input features.

The node degree, defined as the number of edges connected to a node, explicitly encodes structural information about that node's role in the graph. It helps identify ``hub`` nodes~\cite{explainability} that connect to many others and therefore influence a broader neighborhood. Conversely, nodes with lower degree tend to exchange information with fewer neighbors and thus represent more localized and specific features. Thus, including the node degree as an extra feature may help the network modulate the contribution of a node during the message passing phase based on its connectedness or specificity. 

The second feature is the Euclidean distance to the nearest neighbor in the $time \ vs \ wire$ plane, which helps the network distinguish between isolated hits and those that are part of continuous tracks. This ``isolation degree" is particularly relevant for semantic classification and filtering tasks, as expected from physics intuition. For example, minimum ionizing particles (MIPs) and highly ionizing particles (HIPs) form coherent tracks in the $time \ vs \ wire$ plane; thus, isolated hits are unlikely to belong to either category.

Finally, the $\Delta wire$ and $\Delta time$ features also assist in track identification. For each node $i$, the difference is calculated as
\begin{equation}
    \Delta x = 2x_i - x_{j1}-x_{j2}
    \label{eq:deltaX}
\end{equation}
where $x_i$ is the coordinate (either $wire$ or $time$) of the reference node, and $x_{j1}$ and $x_{j2}$ are the coordinates of its two nearest neighbors in the $time \ vs \ wire$ plane, using Euclidean distance. The double differences $\Delta wire$ and $\Delta time$ remain approximately zero along a track (or any segment with fixed direction) as one can see from Figure \ref{fig:edges}. Therefore, this metric can help the network determine whether a node is part of a track, which in turn constrains the likelihood of it belonging to a specific class.

It should be noted that the cost of adding the extra input features is minimal. In the standard NuGraph2, the original four features are first mapped to a vector of size 128 via the encoder. This vector is then concatenated with the original four features, resulting in a 132-dimensional vector before message passing occurs. Adding the four additional features increases the dimensionality only slightly, from 132 to 136.

\begin{figure}
    \centering
    \includegraphics[scale=0.34]{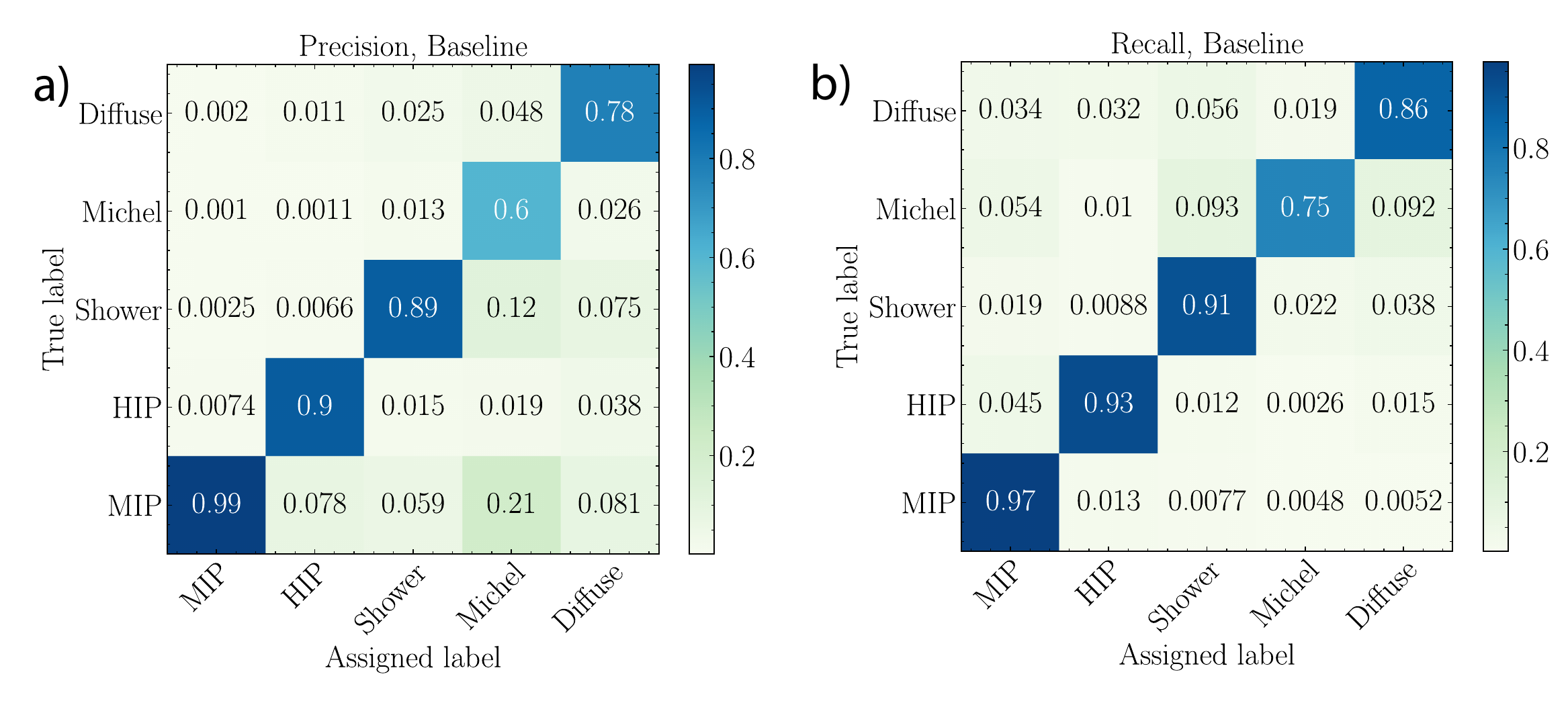}
    \caption{\textbf{a)} Precision and \textbf{b)} Recall confusion matrices for the baseline network.} 
    \label{fig:baseline}
\end{figure}

\begin{figure}
    \centering
    \includegraphics[scale=0.34]{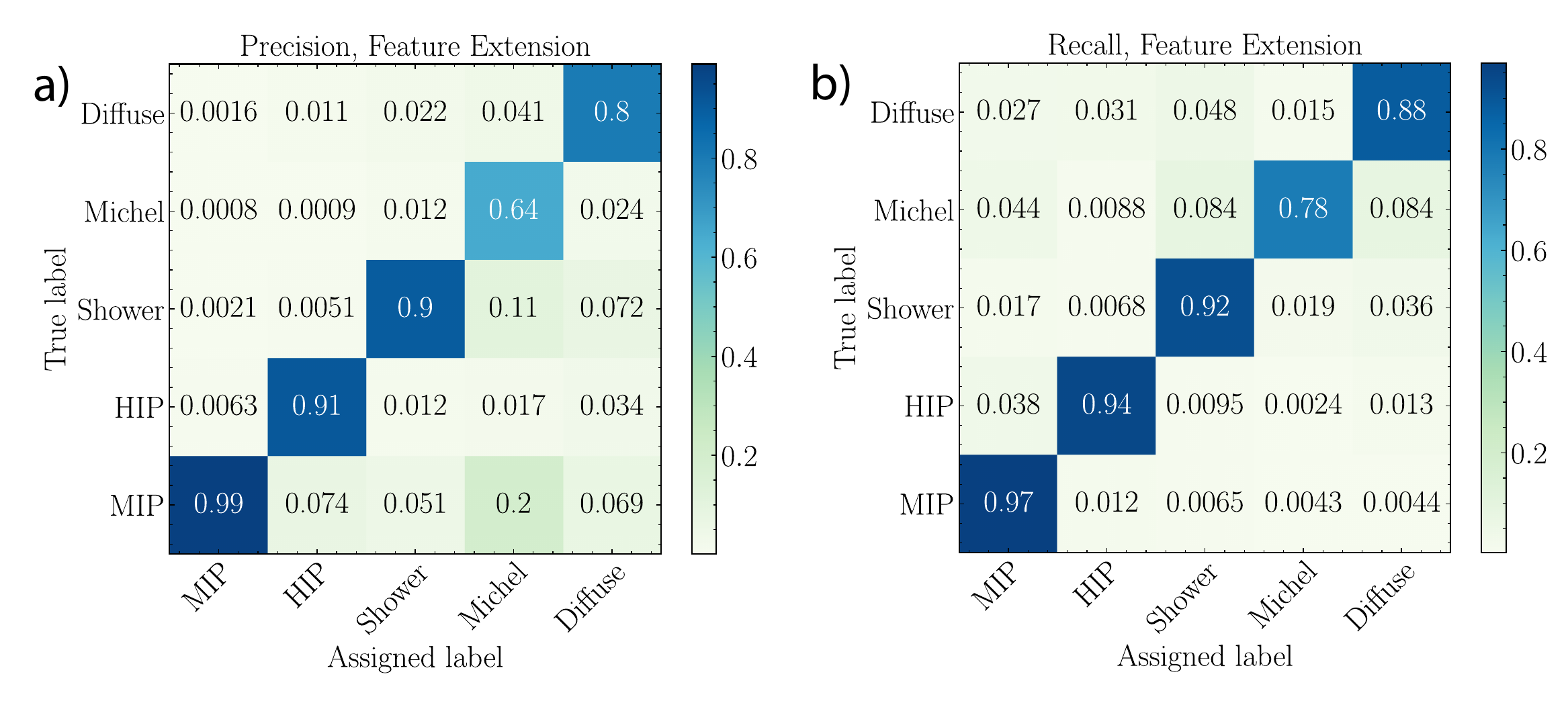}
    \caption{\textbf{a)} Precision and \textbf{b)} Recall confusion matrices for the network with physics-based extended features.} 
    \label{fig:featext}
\end{figure}

Figure~\ref{fig:featext} demonstrates the  classification results obtained by our physics-based extended features as Precision and Recall matrices. Precision is defined as $TP/(TP+FP)$ where $TP$ refers to True Positive and $FP$ to False Positive, and indicates the percentage of particles assigned to class $K$ that actually belongs to $K$. On the other hand, recall is defined as $TP/(TP+FN)$ where $FN$ refers to False Negative, and indicates the percentage of particles from class $K$ that were correctly labeled as $K$. These metrics can also be referred to as purity and efficiency, respectively. Nevertheless, Figure~\ref{fig:featext} shows that both precision and recall metrics are systematically improved when compared to the baseline model's results shown in Figure~\ref{fig:baseline}. The largest gain is observed for Michel electrons, which initially had the worst performance due to their underrepresentation in the dataset. The additional information carried by the new features enabled the network to better extract relevant signals from the few events containing Michel electrons, thereby improving their recognition. An alternative explanation stems from the observation that Michel electrons occupy regions of latent space that overlap with other classes, particularly MIPs. The introduction of additional features provides the semantic head with auxiliary information that helps to better disentangle these overlapping regions in the latent representation space. Overall, this positive result highlights that incorporating context-aware features, crafted using physics intuition, is an effective strategy for injecting domain knowledge and enhancing the network's performance.

While the addition of these features provide improved performance, it may bring an increased risk of domain shift between the training and application datasets. In fact, application datasets which match the training in terms of the nominal (local) input node features, may instead show a discrepancy in terms of the context information added by the new features. For instance, the minimum Euclidean distance to the nearest neighbor may carry information about the global direction of the particle producing the nodes, projected to the specific plane where the nodes are located. The MicroBooNE detector~\cite{acciarri2017design} is exposed to two different neutrino beams entering the detector from different directions. Each beam thus has a preferential direction at which charged particles observed in the detector are produced, i.e. along the neutrino direction. The usage of this feature can thus introduce a domain shift when training on a dataset from one beam and applied to a dataset from the other beam. We also note that the definition of the $\Delta wire$ and $\Delta time$ features is designed to minimize this type of domain shift: in fact the double difference in Eq.~\ref{eq:deltaX} will highlight differences in direction at the node, while a definition based on a single difference, e.g. $\Delta x = x_{j1}-x_{j2}$, would carry information that relates more directly to the global particle direction and may cause domain shift problems.

\subsection{Adding Extra Decoders}
\label{sec:decoders}
In the standard NuGraph2 model, there are two decoders: a semantic decoder and a filter decoder. The semantic decoder classifies particle hits (nodes) into five categories based on the type of particle responsible for the detector hit, while the filter decoder identifies and removes hits that are not associated with the neutrino interaction (e.g., cosmic rays or noise). Both decoders operate on the refined node embeddings produced during message passing. To enhance this shared representation and ultimately improve the semantic task, we introduced an additional decoder designed to predict the distribution of particle classes in an event.

The motivation for this new decoder stems from the relationship between Michel electrons and MIPs. Michel electrons are decay products of muons at rest, and since muons belong to the MIP category, the presence of Michels in an event is conditional on the presence of MIPs. In other words, without MIPs, Michel electrons cannot occur. To exploit this relationship, we designed a decoder to predict the occurrence or distribution of each semantic class within a given neutrino interaction event, with the goal that leveraging class co-occurrence patterns (such as that between MIPs and Michels), would assist semantic segmentation.

The first version was a binary Michel decoder operating at the node level, classifying each node as either Michel (1) or non-Michel (0). The second version shifted to a graph-level Michel counter, returning 1 if any Michel electron was present in the event graph and 0 otherwise. In the third iteration, we expanded the decoder to predict the full class distribution, outputting either the normalized frequency or the percentage of each semantic class per event. Experimental results indicated that percentage-based representations slightly outperformed raw counts. Although this third approach showed improvements over the previous two, the overall accuracy of the semantic classification task remained slightly below the baseline.

A key challenge in implementing this decoder was converting the node feature matrix of shape $(N, K)$, where $N$ varies per event and $K$ is the embedding size, into a fixed-size vector suitable to serve as input to a multilayer perceptron (MLP), while preserving as much information as possible. Our initial attempt used a concatenation of standard pooling operations (sum, max, min, mean), but the resulting vectors lacked sufficient discriminative power: two distinct events could be mapped to similar vectors. To improve this, we implemented a global attention mechanism, where node embeddings were linearly combined using learned attention coefficients. These coefficients were produced by a separate MLP acting on each node embedding and normalized via a softmax function across nodes. This approach enhanced the expressivity of the aggregated representation and increased the standalone accuracy of the count decoder, but it was not sufficient to improve Michel electron metrics beyond the benchmark.

We also experimented with balancing the decoder losses by decreasing the weight of the count decoder relative to the semantic decoder. Although this led to slight improvements in precision and recall, the overall performance still did not surpass the benchmark. Overall, the insights extracted by the network from the auxiliary count decoder task were not sufficient to offset the downside of introducing an additional, competing loss term. The inclusion of this term required the optimizer to balance more tasks, ultimately shifting the optimization trajectory in a way that slightly degraded performance on the targeted semantic task. This means that the gradient directions from the count and semantic loss components did not align very well, causing the overall optimization trajectory to drift away from the optimal minima for the semantic task. 

In summary, incorporating a dedicated Michel electron counter degraded performance. Changing to a general class distribution decoder offered modest improvements with respect to the Michel electron counter, but still fell short of outperforming the baseline. Although the count decoder introduced auxiliary supervision, the representational gains it provided were insufficient to compensate for the additional optimization burden imposed by its loss term. 
It is worth noting that this decoder is expected to become more effective under a more granular particle labeling scheme. At present, many distinct particles fall under the broad HIP and MIP categories. Future iterations of NuGraph are anticipated to incorporate a richer set of particle labels, enabling the count decoder to capture more meaningful inter-class relationships. These relationships can in turn support the construction of a more expressive internal representation of each hit and therefore enhance the performance of the semantic decoder.

\subsection{Michel Energy Regularization}
\label{sec:energyreg}
Michel electrons are the product of the decay of muons at rest in the detector. Their energy spectrum is governed by energy conservation laws applied to the kinematics of the three-body decay process~\cite{ParticleDataGroup:2024cfk}.
To improve the Michel electron metrics, we introduced a physics-based regularization term in the loss function that penalizes the model for classifying nodes as Michel electrons when the sum of the energies deposited by these nodes exceeds a threshold or deviates from the expected energy distribution observed in the batch. The key idea is that while a single Michel electron can produce a variable number of detector hits in each LArTPC plane, the total deposited energy across those hits is ultimately constrained by the energy of the originating Michel electron~\cite{MicroBooNE:2017kvv}. We tested three different regularization schemes:

\begin{itemize}
\item Penalization based on how much the Michel energy exceeds a fixed energy cutoff.
\item Penalization based on deviation from the expected Langauss distribution.
\item Penalization based on deviation from the expected distribution derived from simulation data.
\end{itemize}

The first scheme penalizes the network for assigning the Michel label to nodes whose deposited energy exceeds the expected range for Michel electrons, as observed in the dataset. Since hits in our dataset do not measure the deposited energy directly, we used one of their features as a proxy, i.e. the waveform integral ($Integral$). 
This correlation is modeled through a chain of different tools. The propagation of particles in the detector volume is simulated with Geant4~\cite{GEANT4:2002zbu} v10, which produces a collection of true energy deposits in the detector volume. The corresponding ionization charge is then propagated in the detector using various tools available in LArSoft~\cite{Snider:2017wjd}. The recombination of electron-ion pairs is modeled with a modified box model~\cite{ArgoNeuT:2013kpa}. Distortions in the electric field due to the space-charge effects are simulated with data-driven electric field maps~\cite{MicroBooNE:2019koz, MicroBooNE:2020kca}. Other non-uniformities, resulting from variations in electron lifetime or wire response are also simulated. The readout electronics and wire response are modeled through a simulation of the charge induced by electrons that drift to the wire planes at the anode ~\cite{MicroBooNE:2018swd, MicroBooNE:2018vro}. Waveforms are then processed through a reconstruction chain that includes noise filtering~\cite{MicroBooNE:2017qiu}, signal deconvolution and region-of-interest finding~\cite{MicroBooNE:2018swd}. Hits~\cite{Baller:2017ugz, Berkman:2021ffy} are Gaussian fits to peaks in regions of interest on the deconvolved waveform. Their Integral, expressed in ADC counts, is computed from the corresponding best-fit Gaussian function parameters. In addition to the instrumental effects listed above, deviations from an ideal linear dependence can also be due to other contributions to the hit integral in addition to the true Michel energy deposition. In particular, other particles, such as cosmics, with trajectories overlapping with Michel hits in the wire readout plane can lead to a substantial bias towards larger integral values.

\begin{figure}
    \centering
    \includegraphics[scale=0.4]{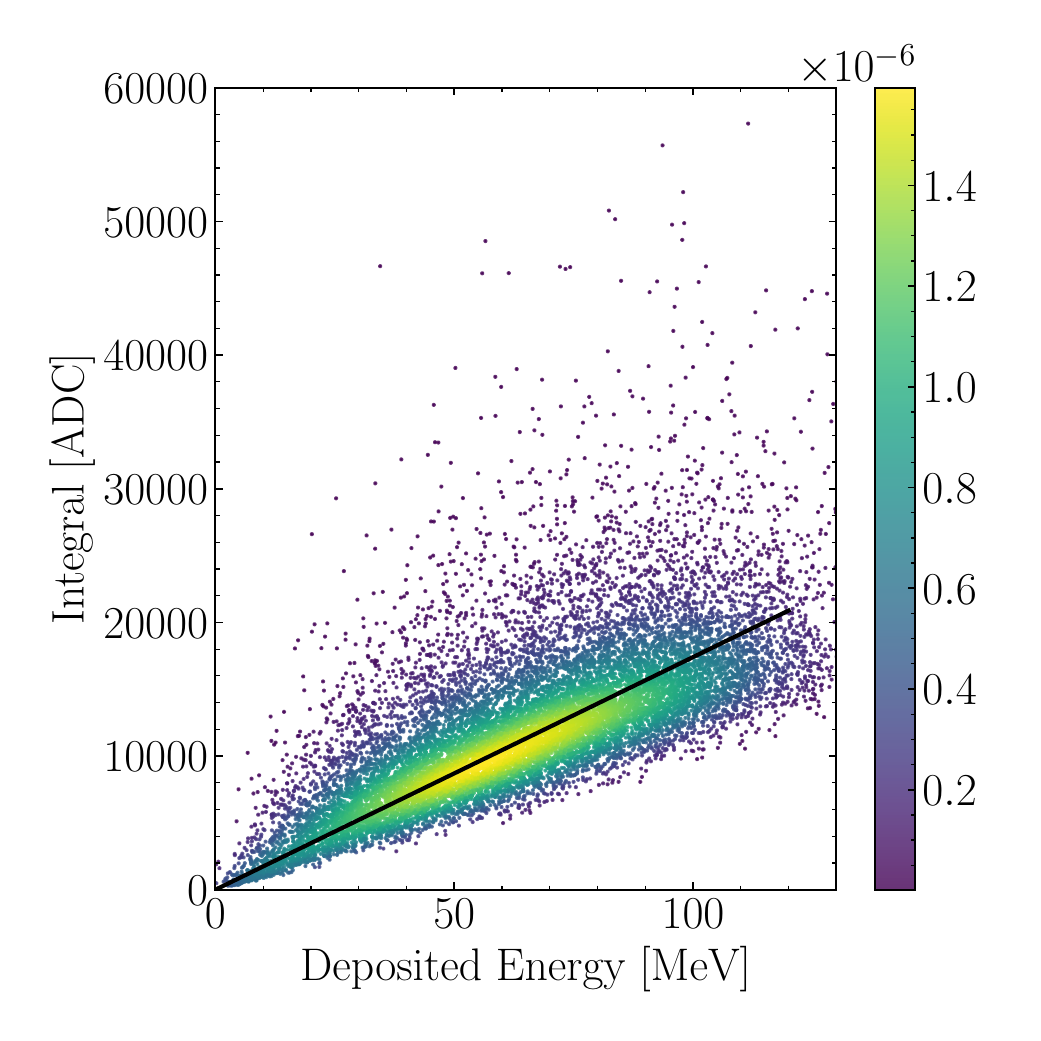}
    \caption{Distribution of Michel electrons in the $Integral$ $vs$ Deposited Energy summed across the three planes (event-wise). The reported $Integral$ is the sum of the integral of all hits with Michel as their true label in each event across all LArTPC planes, while Deposited Energy is the sum of the corresponding simulated energy deposits.} 
    \label{fig:kde}
\end{figure}

Figure~\ref{fig:kde} shows the distribution of Michel electron hit integral vs. deposited energy; as expected, they are approximately linearly related, with the slope of the fitted black line encoding this relationship. This allowed us to estimate the deposited energy from the $Integral$ hit feature and apply the regularization. This regularization is expressed as a penalty term to the loss function:
\begin{equation}
    \mathcal{L}_{MichReg} = c_1(E - E_{cut}) \ \ \mathrm{if} \ E>E_{cut}
\label{eqn:reg}
\end{equation}
where $c_1$ is the regularization weight, $E$ is the node's predicted deposited energy, and $E_{cut}$ is a hyperparameter defining the energy threshold beyond which the penalty is applied. The value of $E_{cut}$ was chosen based on the distribution in Figure~\ref{fig:kde}, and we tested a range of values between 90 and 160 MeV. This loss term was evaluated whenever the semantic decoder produced its predictions. If Michel electrons were present in the event, their deposited energies were computed, and Eq.~\ref{eqn:reg} was applied. 

\begin{figure}
    \centering
    \includegraphics[scale=0.4]{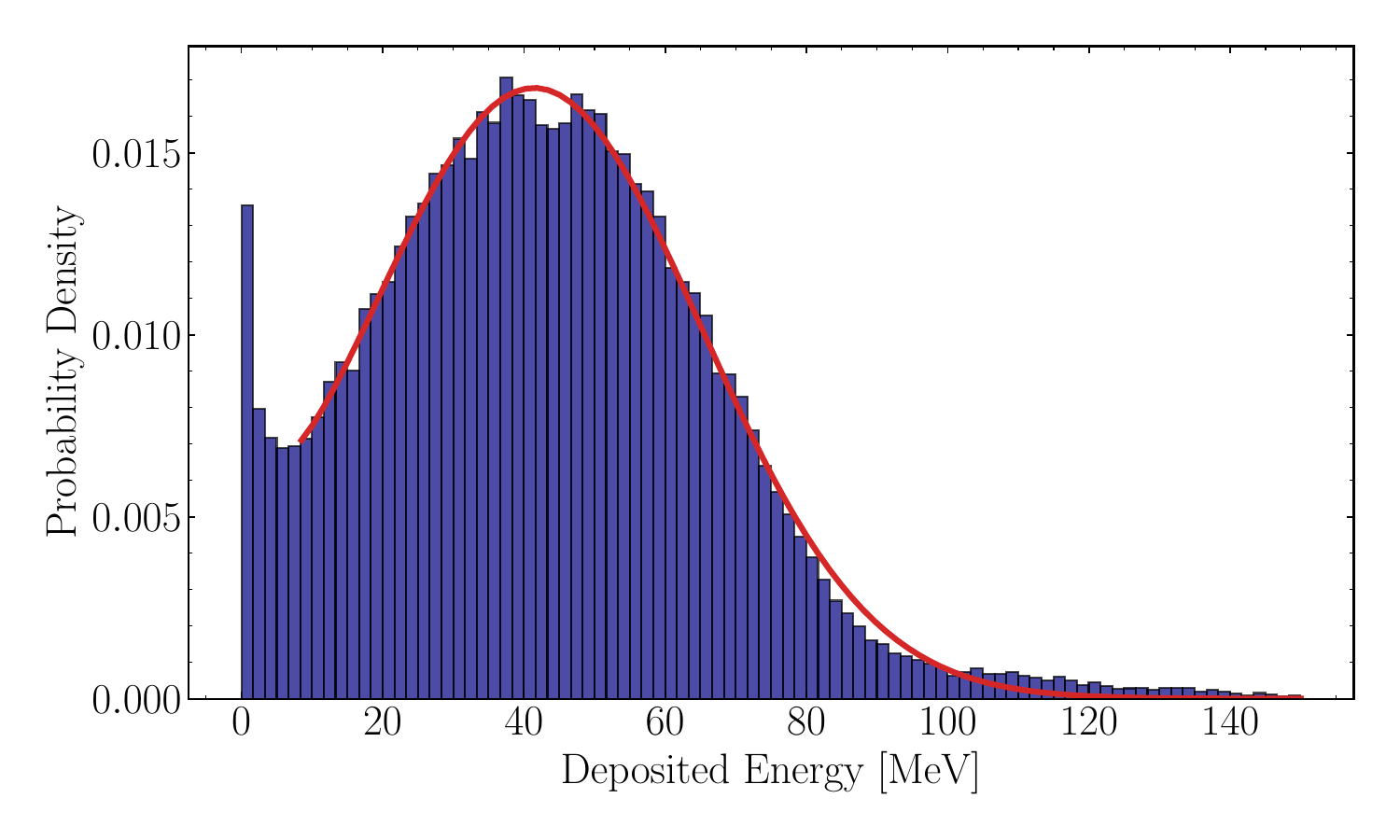}
    \caption{Michel electron deposited energy distribution summed across all planes (i.e., for the entire event). The red curve indicates the Langauss fit, starting at $E=8.5$ MeV. See Fig.~\ref{fig:kde} for a description of how the Deposited Energy is computed.} 
    \label{fig:fit}
\end{figure}

We also explored an alternative regularization scheme based on how well the observed Michel energy distribution matches predictions from our simulation. Figure~\ref{fig:fit} shows the expected deposited energy spectrum of Michel electrons, summing contributions across all LArTPC planes. The peak near $E = 0$ MeV is due to Michel electrons propagating mostly outside the detector fiducial volume or in regions dominated by unresponsive wires. We use two methods to model the expected Michel energy. In the first method, we aim at modeling the probability based on a functional form that accounts for cases where the Michel deposits most of its energy in the detector. We therefore fit the histogram in Figure~\ref{fig:fit} with a Langauss fit (red curve), computed only for $E \geq 8.5$ MeV. In contrast, the second method uses the actual histogram in Figure~\ref{fig:fit} to define a probability density function that includes all detector effects, including the peak at 0 MeV. We penalized the model for differences between the Michel energy distribution obtained from the NuGraph predictions and the expected distribution from the simulation, computed either with the Langauss fit or using the empirical simulation distribution.

It is important to note that including Michel-specific regularization terms can inadvertently discourage the network from labeling nodes as Michel electrons, particularly if the regularization loss becomes comparable to the semantic loss. To mitigate this, we ensured that the regularization weight was significantly smaller than that of the semantic loss. Despite this precaution, neither regularization scheme improved Michel electron classification metrics. Even with a small regularization weight, the network became overly cautious, resulting in decreased recall and accuracy for Michel predictions.

Two other limitations in this approach that likely contribute to the degradation in performance are the following. The first is related to the relationship between the $Integral$ feature and deposited energy. Although Figure~\ref{fig:kde} shows a clear positive linear trend for Michel hits, the large spread of data points around the fitted line indicates significant variance. This implies that while $Integral$ and energy are correlated, the uncertainty in this relationship limits the reliability of energy estimation based solely on $Integral$, thereby undermining the effectiveness of energy-based regularization. A similar behavior is observed in Figure~\ref{fig:fit}, where the Langauss fit, represented by the red line, fails to accurately capture the fast decay and extended tail of the data distribution above the 80 MeV region. However, note that most of the probability mass is concentrated below that region. Thus, although the fit is less accurate in the region beyond 80 MeV, it is still adequate for a proof-of-principle, and the residual mismatch has little influence on our conclusions.
The second limitation relies in the fact that NuGraph2 only makes predictions about hits, and does not have a concept of a particle instance with a given deposited energy. Consequently, there is not a one-to-one correspondence between a set of nodes semantically labeled according to a given class and a specific particle unless there is only a single particle of that class in the event. This means that the penalty term is computed for each event in the batch, which represents a coarser granularity than a more natural per-particle contribution. Improving these limitation is beyond the scope of this paper, as it will require completion of ongoing updates to NuGraph that will allow to predict per-particle attributes.

\section{Discussion}
Tables \ref{tbl:prec} and \ref{tbl:rec} summarizes the results of all three interventions and the baseline network. The inclusion of extra input features, designed with physics intuition, proved to be the most effective intervention for our application. These features encoded non-local and structural information, such as node degree, local geometric context in the $time \ vs \ wire$ plane, and differential measures that reflect local track continuity, which significantly improved semantic classification metrics across all classes. In particular, Michel electron recall and precision showed the largest gains due to their previously poor performance caused by class imbalance and latent space overlap. This result highlights the value of embedding physics-informed context directly into the model's input representation, as it helps disentangle class boundaries in the learned latent space.

In contrast, the use of additional decoders to supervise graph-level class distributions showed limited success. Although the multi-class count decoder yielded modest improvements over simpler Michel-specific counters, none of these approaches surpassed the baseline performance. These auxiliary tasks introduced competing gradients in the optimization process, subtly degrading the model's ability to fine-tune the semantic classification head.

Similarly, physics-based regularization terms targeting Michel electron energy distributions failed to improve classification performance. Despite careful calibration of regularization strength and theoretical grounding in energy expectations, these terms caused the network to become overly conservative in assigning the Michel label. This result was attributed in part to the high variance in the relationship between waveform integral and true deposited energy, which weakened the reliability of energy-based penalties. This challenge is compounded by the lack of a one-to-one mapping between a set of graph nodes and physical particles (a current limitation of NuGraph2), which blurs the link between node-level energy and the true energy of the originating particle.

It is worth noting that both the use of auxiliary decoders and the incorporation of physics-inspired loss terms remain broadly valid strategies for injecting domain knowledge into deep learning models. Their limited effectiveness in our case may stem from specific characteristics and limitations of the NuGraph2 architecture, the features of our dataset, or the particularities of Michel electron identification. NuGraph2 is limited to hit-level decoders and therefore lacks internal representations at the particle or event level. As a result, both the count decoder and the Michel energy regularization, which are both inherently more meaningful at these higher levels, do not integrate well with the current architecture. These approaches are expected to align more naturally with the forthcoming hierarchical NuGraph3, which will explicitly incorporate particle- and event-level information. We therefore view these approaches as promising in principle, and potentially effective in other architectures or problem settings.

\begin{table}[h!]
  \caption{Semantic decoder precision for all particle labels for baseline and all three interventions.}
  \label{tbl:prec}
  \begin{tabular}{llllll}
    \\
    \hline
    Technique & MIP & HIP & Shower & Michel & Diffuse  
    \\
    \hline
    Baseline & 0.99 & 0.90 & 0.89 & 0.60 & 0.78\\ 
    Feat. Ext. & \textbf{0.99} & \textbf{0.91} & \textbf{0.90} & \textbf{0.64} & \textbf{0.80}\\ 
    Count Decoder & 0.98 & 0.89 & 0.86 & 0.48 & 0.75\\
    Michel Reg. & 0.99 & 0.90 & 0.88 & 0.58 & 0.78\\
    \hline
  \end{tabular}
\end{table}

\begin{table}[h!]
  \caption{Semantic decoder recall for all particle labels for baseline and all three interventions.}
  \label{tbl:rec}
  \begin{tabular}{llllll}
    \\
    \hline
    Technique & MIP & HIP & Shower & Michel & Diffuse  
    \\
    \hline
    Baseline & 0.97 & 0.93 & 0.91 & 0.75 & 0.86\\ 
    Feat. Ext. & \textbf{0.97} & \textbf{0.94} & \textbf{0.92} & \textbf{0.78} & \textbf{0.88}\\ 
    Count Decoder & 0.96 & 0.91 & 0.89 & 0.68 & 0.83\\
    Michel Reg. & 0.97 & 0.93 & 0.91 & 0.74 & 0.86\\
    \hline
  \end{tabular}
\end{table}

\section{Conclusions}
In this work, we explored multiple strategies to improve semantic classification in the NuGraph2 architecture by injecting physics-informed inductive biases into the model. We focused on three primary directions: enriching the input feature space with physics-inspired quantities, introducing auxiliary decoders that capture class-level structure, and incorporating energy-based regularization terms derived from physical expectations.

Overall, our findings suggest that targeting Michel electrons in isolation is ineffective; progress arises when the latent representation is sharpened for all classes, allowing the minority class to benefit indirectly. Enhancing the model's understanding of global class relationships and input structure benefits all classes, including Michels. For the NuGraph2 application, our results show the importance of integrating domain knowledge at the representation level, rather than solely through auxiliary objectives or task-specific losses. 

All the physics-injection approaches we explored are generally valid, and their performance may lead to different level of improvements on other applications, including the hierarchical version of NuGraph that is currently under development.

\section{Acknowledgments}
The authors would like to thank Burt Holtzman and Maria Acosta for their management of computing resources used for running a large amount of the analysis included in this work. 
We acknowledge the MicroBooNE Collaboration for making publicly available the datasets~\cite{abratenko_2022_7261921, abratenko_2023_8370883} employed in this work. 

This work was produced by FermiForward Discovery Group, LLC under Contract No. 89243024CSC000002 with the U.S. Department of Energy, Office of Science, Office of High Energy Physics. Publisher acknowledges the U.S. Government license to provide public access under the DOE Public Access Plan DOE - \url{http://energy.gov/downloads/doe-public-access-plan}. This work was funded by a grant from the University of Illinois Discovery Partners Institute.

\bibliographystyle{IEEEtran}
\bibliography{cite}
\end{document}

%% file: packages.tex
\usepackage[ascii]{inputenc}
\usepackage[T1]{fontenc}
\usepackage{geometry}

\usepackage{amsmath}
\usepackage{amssymb}
\usepackage[numbers,sort&compress]{natbib}
\usepackage[compact]{titlesec}
\usepackage{floatflt}
\usepackage{longtable}
\usepackage{wrapfig}
\usepackage{floatrow}
\usepackage{graphicx}
\usepackage{tabularx}
\usepackage{subcaption}
\usepackage{url}
\usepackage[usenames,dvipsnames]{color}
\usepackage{pdfpages}
\usepackage{enumitem}
\usepackage{authblk}

\usepackage{mathptmx}
\usepackage[english]{babel}
\usepackage{amsmath}
\usepackage{amssymb,amsfonts,textcomp}
\usepackage{array}
\usepackage{supertabular}
\usepackage{multirow}
\usepackage{hhline}
\usepackage{enumitem}
\usepackage{xspace}
\usepackage{upgreek}
\usepackage{microtype}
\usepackage{lineno}
\usepackage{stackrel}
\usepackage{accents}

%% file: authors.tex
\author{%
    Vitor F. Grizzi$^{1,2}$\footnote{\texttt{vferreiragrizzi@anl.gov}}, 
    M. Voetberg$^3$, 
    V Hewes$^4$,
    Giuseppe Cerati$^3$, 
    Hadi Meidani$^5$\\

    \small{
    $^1$ Department of Nuclear, Plasma, and Radiological Engineering, Grainger College of Engineering, University of Illinois Urbana-Champaign, Urbana, IL 61801
    
    $^2$ Chemical Sciences and Engineering Division, Argonne National Laboratory, Lemont, IL 60439
    
    $^3$Fermi National Accelerator Laboratory,  Batavia, IL 60510, USA
    
    $^4$ Department of Physics,  University of Cincinnati, Cincinnati, OH 45221-0377
    
    $^5$ Department of Civil and Environmental Engineering,  Grainger College of Engineering, University of Illinois Urbana-Champaign, Urbana, IL 61801

    }
}